\documentclass[preprint,11pt]{article}

\usepackage{bm}
\usepackage{epsfig}
\usepackage{color}
\usepackage{amssymb}
\usepackage{amsmath}

\newcommand {\lab}[1]{\label{eq:#1}}

\newcommand {\be}[1]{\begin{equation}{\lab{#1}}}
\newcommand {\ee}{\end{equation}}
\newcommand {\bea}{\begin{eqnarray}}
\newcommand {\eea}{\end{eqnarray}}

\begin{document}

\title{Numerical integration of variational equations for Hamiltonian systems with long range interactions}

\author{
\textbf{Helen Christodoulidi$^{1}$, Tassos Bountis$^{1}$ and Lambros Drossos$^{2}$}\\
$^{1}$Department of Mathematics, Division of Applied Analysis  
and \\ Center for Research and Applications of Nonlinear 
Systems (CRANS),\\ University of Patras, GR-26500 Patras, Greece. \\
$^2$High Performance Computing Systems and 
Distance Learning Lab (HPCS-DL Lab),\\
Technological Educational Institute of Western Greece,\\ GR-26334 Patras, Greece}


\maketitle

\begin{abstract}
We study numerically classical 1-dimensional Hamiltonian lattices involving inter--particle long range interactions that decay with distance like $1/r^{\alpha }$, for $\alpha \geq 0$. We demonstrate that although such systems are generally characterized by strong chaos, they exhibit an unexpectedly organized behavior when the exponent $\alpha < 1$. This is shown by computing dynamical quantities such as the maximal Lyapunov exponent, which decreases as the number of degrees of freedom increases. We also discuss our numerical methods of symplectic integration implemented for the solution of the equations of motion together with their associated variational equations. The validity of our numerical simulations is estimated by showing that the total energy of the system is conserved within an accuracy of 4 digits (with integration step $\tau =0.02$), even for as many as $N=8000$ particles and integration times as long as $10^6$ units.
\end{abstract}

                             
 
 \section{Introduction}
\label{intro}
Hamiltonian systems describing 1-Dimensional (1D) particle chains characterized by interactions of various ranges constitute an active area of research with increasing interest due to their applicability in many scientific fields. In particular, the relevance of long (vs. short) range interactions has been extensively studied and intensely debated in a wide variety of problems of statistical mechanics, mean field theories, active matter, dynamical networks, etc. regarding the various degrees of chaos involved in their time evolution. In statistical physics for example, the classical Boltzmann framework for the appropriate entropy functional at thermal equilibrium is not adequate for describing systems with long range interactions, as remarked already by J. W. Gibbs \cite{Gibbs1902}. 

In the past 25 years, a great number of researchers \cite{Tsallis1988,GellMannTsallis2004,PluchinoRapisardaTsallis,
Tsallis2009,Tsallis2014,Antonopetal2011} have shown that there exist long lasting quasi--stationary states (QSS) in a variety of physical and biological systems characterized as {\em non--additive}, i.e. that cannot be decomposed in entirely independent parts. In such cases, a different entropy functional (the so--called Tsallis entropy) appears to be more suitable for their thermodynamic description, while the associated probability distribution functions are of the $q$--Gaussian type with $q>1$. This divergence from the classical Boltzmann--Gibbs Maxwellian distributions of $q=1$, raises new questions regarding the statistical and dynamical behavior of such systems in the thermodynamic limit of very large $N$ and total energy $E$, with $E/N$ constant.

As we have demonstrated in recent publications \cite{CWAB,CTB}, systems with long range interactions (LRI) have significant advantages, since they often exhibit {\em a weaker chaos} than those with interactions only between nearest neighbors. For example, in active matter systems consisting of self-propelled particles (like birds) it has been observed that nonlocal communication acts as a counterbalance against external threats or attacks. This is the case, for instance with flocks of starlings described by the so--called topological model introduced by the STARFLAG group \cite{parisi}. Their observations on groups of starlings in Rome revealed that synchronized movements are based on a fixed number of interacting neighbors, independent of the distance between them. Moreover, it was shown in \cite{CWAB} that the number $n$ of interacting particles in the topological model is crucial for the coherence of the group: $n$ needs to be large enough to overcome random perturbations as well as maintain cohesion.

On the other hand, the study of LRI in the classical framework of dynamical systems is of great interest. Spatially localized oscillations called breathers, well--known as simple periodic solutions of lattices with nearest neighbor interactions, make their appearance also in Hamiltonian systems with LRI (see \cite{Flach98} for more details). Another interesting type of collective behavior occurs in the form of long-living QSS of the type mentioned above, while
in a system of $N$ coupled planar rotators that interact via long-range forces 
critical regions were found where these states appear \cite{latora}. Moreover, in \cite{latora,AnteneodoTsallis1998} a systematic study of the largest Lyapunov exponent showed that it decays as a power-law with $N$, making the system quasi--integrable in the thermodynamic limit. Such QSS were also studied in a generalized mean field system, where interactions decay with distance according to $1/r^{\alpha }$ \cite{campa,CirtoAssisTsallis2013}. 

More recently, the Fermi-Pasta-Ulam-$\beta $ model (FPU-$\beta $) with LRI was studied in \cite{CTB,Ruffo14} and opened a new branch of research in this field. In \cite{Ruffo14} for example, the authors explore instability regimes for the low frequency modes of an FPU-$\beta $ chain in relation to the long--standing question of the relaxation times required to reach energy equipartition among all modes.

In the present paper we focus on two topics: (i) First we extend recent studies of nearest-neighbor Hamiltonian lattices to analogous models involving LRI, and (ii) we employ numerical integration schemes to compute the tangent dynamics needed for the calculation of the largest Lyapunov exponent. The structure of our paper is as follows:  In Section \ref{LRI} we discuss in detail the transition to LRI for different boundary conditions, while in Section \ref{VE} we solve the associated variational equations for specific chaotic states. Subsection \ref{2si} gives the basic properties and definitions of symplectic integrators and discusses in Subsection \ref{tmm} the so--called tangent map method. Finally, in Section \ref{NR} we apply these techniques to two different Hamiltonian systems: the mean field model of planar rotors and the FPU-$\beta $ chain, both with interactions modulated by the factor $1/r^{\alpha }$. We end with our conclusions in Section \ref{concl}.


\section{Hamiltonian particle chains with long range interactions}
\label{LRI}

Let us consider a 1D Hamiltonian chain with quadratic kinetic energy in the generalized momenta
and a potential that depends purely on the generalized positions and contains nearest neighbor interactions together with an on-site potential. This system is described by the Hamiltonian
\begin{eqnarray}\label{hamgen}
{\cal H}=\frac{1}{2}\sum_{i} p_i^2 + \sum_{i}W(x_{i}) + \sum_{i}V(x_{i+1}-x_{i}) ~~,
\end{eqnarray}
where $p_i$ and $x _i$ are canonical conjugate pairs of positions and momenta and boundary conditions are chosen arbitrarily.

\begin{figure}
\centering
\includegraphics[width=0.35\linewidth]{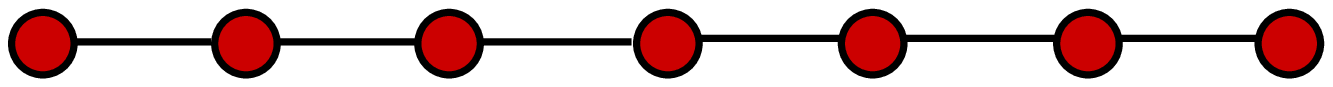}
\includegraphics[width=0.35\linewidth]{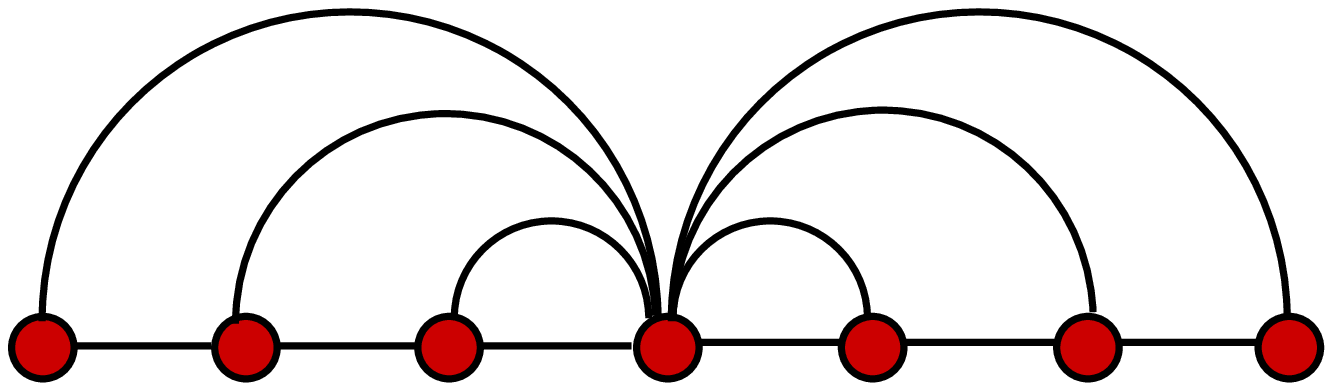}\\
\includegraphics[width=0.2\linewidth]{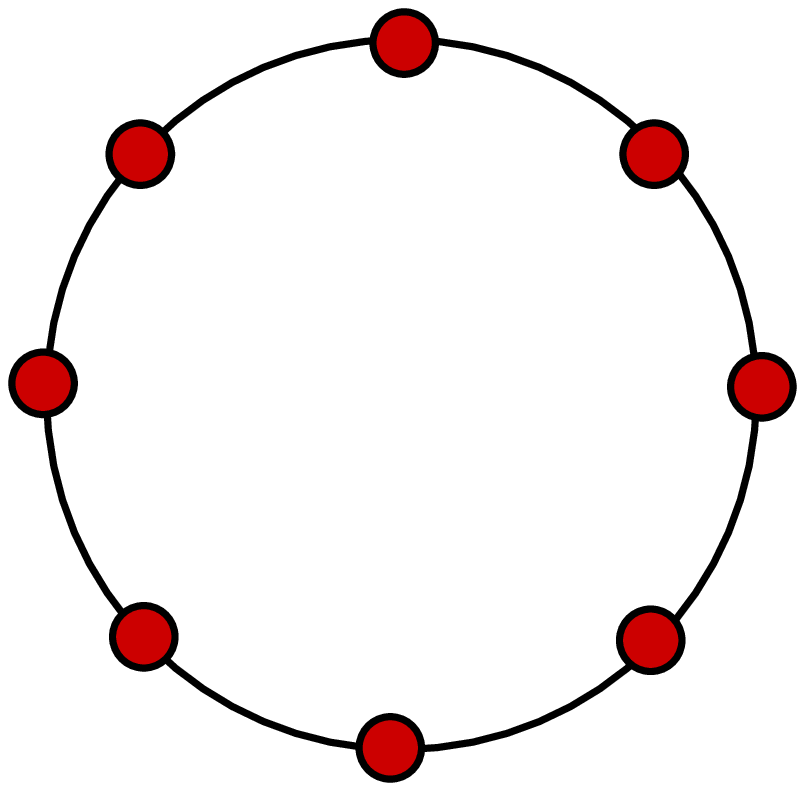}
\includegraphics[width=0.22\linewidth]{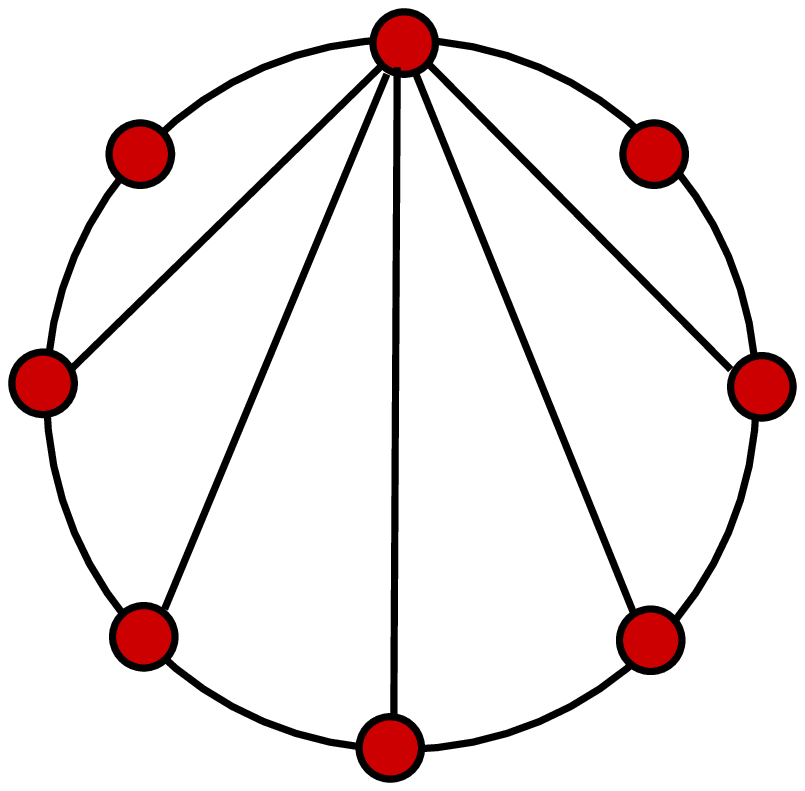}
\caption{ Lattices with nearest neighbor interactions ($\alpha \rightarrow \infty$) on the left are converted into long range systems on the right. In our paper the strength of this interaction decays with distance as $1/r^{\alpha }$, for finite $\alpha \geq 0$. Upper panels correspond to fixed or open boundary conditions and lower panels to periodic boundary conditions.} \label{transition}
\end{figure}

To convert the 1D lattice (\ref{hamgen}) to an LRI system, all nearest neighbor interacting terms\footnote{Or some of them. In \cite{CTB} we considered LRI on the quartic potential of FPU-$\beta $ model, while the quadratic one was left unchanged.} in the potential, i.e the position differences $x_{i+1}-x_i$, should be replaced by the difference combinations
$x_i-x_j$, $i,j=1,\ldots,N$, so that particles can form a fully connected graph (see Fig. \ref{transition}). Then, of course, one can apply interactions which depend on the topological distance and decay with distance as $1/r^{\alpha }$, with $\alpha \geq  0 $, as in \cite{AnteneodoTsallis1998}. Next, (\ref{hamgen}) is converted to a Hamiltonian with LRI:

\begin{eqnarray}\label{hamlri}
{\cal H}=\frac{1}{2}\sum_{i} p_i^2 + \sum_{i}W(x_{i}) + \frac{1}{2 \tilde{N}}\sum_{\mathop{i,j}\limits_{i\neq j}} \frac{V(x_{i}-x_{j})}{r_{i,j}^{\alpha }} ~~,
\end{eqnarray} 

where $r_{i,j}$ defines the topological distance, i.e. the minimum connection
between the particles $i$ and $j$. $\widetilde N$ is the rescaling factor necessary for making the Hamiltonian (\ref{hamlri}) extensive, which means that the potential function $V$ is proportional to the system size\footnote{Otherwise the sum with long range terms would be approximately proportional to $N~ ^2$ and
would converge faster to infinity than the other two sums of (\ref{hamlri}) in the thermodynamic limit.}. In particular, when the chain has fixed or open boundary conditions, we set
\begin{eqnarray}\label{factor}
r_{i,j}=\mid i-j\mid  ~~{\mathrm {and}}~~{\widetilde N}  = \frac{1}{N} \sum_{i=0}^{N} \sum_{j=i+1}^{N+1}  \frac{1}{(j-i)^\alpha}~~,
\end{eqnarray}
and when it has periodic boundary conditions we write:
\begin{eqnarray}\label{factor}
r_{i,j}={\mathrm {min}} \{ \mid i-j\mid, N- \mid i-j \mid\}~~{\mathrm {and}}~~ {\widetilde N}  = 2 \sum_{i=1}^{N/2-1}\frac{1}{i^{~\alpha}}~~. 
\end{eqnarray}

The role of the parameter $\alpha $ which controls the range of interaction is essential. The two extreme cases are:
(i) $\alpha \rightarrow \infty $, where only interactions between first neighbors apply and (ii) $\alpha =0$, 
which corresponds to the mean field model (HMF), where all particles interact equally and independently of 
their distance, as in a fully connected graph.

It appears that the implementation of LRI in a Hamiltonian particle chain strongly affects the system's dynamics and statistics. The excitation of a single site results in an almost immediate reaction of all the particles. The reaction time is significantly longer in simple chains. There is a number of physically interesting models whose dynamics can be extended to long range and thoroughly studied to provide representative examples for the LRI effect: (i) the Klein-Gordon chain for $W(x)=\frac{1}{2}x^2 \pm  \frac{K}{4}x^4$ and $V(x)= \frac{\varepsilon}{2} x^2$ in (\ref{hamlri}), (ii) a system consisting of coupled planar rotators with $V(x)={1-\cos x}$ (and $W(x)=0$) and (iii) the Fermi-Pasta-Ulam-$\beta $ (FPU-$\beta $) model with $V(x)=\frac{1}{2}x^2+ \frac{\beta }{4}x^4$ in (\ref{hamlri}). Systems (ii) and (iii) are the ones treated in the present paper, while extensions of the KG model (i) will be dealt with in a future publication.

 
\section{Numerical integration of LRI systems and their tangent dynamics}
\label{VE}
The equations of motion for the Hamiltonian system (\ref{hamlri}) with LRI read:
\begin{eqnarray}\label{hameq}
\dot{x_i} &=& p_i \nonumber\\
\dot{p_i} &=& - \frac{\partial W(x_i)}{\partial x_i} - \frac{1}{ \tilde{N}}\sum_{\mathop{j}\limits_{j\neq i}} \frac{1}{r_{i,j}^{\alpha }} \frac{ \partial V(x_{i}-x_{j})}{\partial x_i} ~~.
\end{eqnarray}

The tangent dynamics (or variational equations) of a dynamical system is of great importance in uncovering the local properties of its solutions that lead to the estimation of chaotic indicators such as Lyapunov and GALI exponents \cite{GALI}. In the case of a mechanical system
the variational equations are defined on the tangent space { of the configuration space, constituting a 
4$N$ dimensional phase space. In particular, the variational equations of the system (\ref{hamlri}) can be written in the following form:}
\begin{eqnarray}\label{vandw}
 \left(
\begin{array} {ll}
{\bf \dot{\delta x} }\\
{\bf \dot{\delta p}}
\end{array}
\right) = \left(
\begin{array} {ll}
O_N  &  I_N  \\
A_N &  O_N\\
\end{array}
\right) \cdot \left(
\begin{array} {ll}
{\bf \delta x}\\
{\bf \delta p}
\end{array}
\right)  ~~,
\end{eqnarray}
where each element of the matrix $A_N$ is given by the second partial derivatives 
\begin{equation}
a_{i,j}= \left\{
\begin{array} {ll}
-\frac{\partial ^2 W(x_i)}{\partial x_i^2} -\frac{1}{ \tilde{N}}\sum_{\mathop{k}\limits_{i\neq k}} \frac{1}{r_{i,k}^{\alpha }} 
\frac{ \partial^2 V(x_{i}-x_{k})}{\partial x_i^2}, ~~~ {if}~~i=j\\
- \frac{1}{ \tilde{N}} \frac{1}{r_{i,j}^{\alpha }} \frac{ \partial^2 V(x_{i}-x_{j})}{\partial x_i \partial x_j}, ~~~~~~~~~~~~~~~~~~~~~~~~~{if}~~i\ne j
\end{array}
\right.
\end{equation}
calculated along the orbit $({\bf x}(t),{\bf p}(t))$, while $O_N,I_N$ are the zero and identity $N \times N$ matrices respectively.

Now the equations of motion (\ref{hameq}) together with their variations (\ref{vandw})
constitute a time-dependent Hamiltonian system defined on the tangent space of the system (\ref{hamlri}) and expressed by:
\begin{eqnarray}\label{hamvar2}
{\cal H}=\frac{1}{2}\sum_{i} \delta p_i^2 + \frac{1}{2}\sum_{i} \frac{\partial ^2 W(x_i)}{\partial x_i^2} (\delta x_i) ^2
+ \frac{1}{ 2 \tilde{N}}\sum_{\mathop{i,j}\limits_{i\neq j}} \frac{1}{r_{i,j}^{\alpha }} \frac{ \partial^2 V(x_{i}-x_{j})}{\partial x_i \partial x_j}
 \delta x_i \delta x_j~~.
\end{eqnarray}

\subsection{Symplectic integrators}
\label{2si}

Symplectic integration schemes are designed to conserve the symplectic structure of a Hamiltonian system exactly, but not its total energy. During the computations, neither the orbit  nor the Hamiltonian $H$ itself are precisely followed. Even though these are not desirable properties for an integration method, symplectic integration  still has great advantages compared with other methods \cite{yoshida}:
(i) There exists an analytic autonomous Hamiltonian system $\tilde{H}$ which is followed exactly \cite{benettin94} and provides the so--called backward error analysis.
(ii) Energy errors are bounded and hence do not increase for exponentially long times.

For reasons of self--consistency and in order to help the reader who is not familiar with symplectic integration theory, we give here some notations in the spirit of Lie algebras \cite{neri,yoshida}. 
A symplectic integrator approximates the solution $\vec{x}(t) = e^{t L_H \vec{x}(0) }$ of the system by splitting 
the kinetic and potential terms of the Hamiltonian, i.e. $e^{\tau  L_H  } = e^{\tau (A+B) } $,
where $A:=D_T$ is the kinetic and $B:=D_V$  the potential operator, which do not commute in general.
If we denote by $\tau $ the time step of the algorithm, the truncation in the Taylor expansion of 
$e^{\tau  L_H  } $ determines the order of precision in terms of $\tau $.  

The standard 2nd order symmetric exponential splitting \cite{yoshida}, is:
\begin{eqnarray}
S_2(\tau (A+B)):= e^{\frac{1}{2}\tau A} e^{ \tau B} e^{\frac{1}{2}\tau A}+ {\cal O}(\tau ^3) \nonumber ~~,
\end{eqnarray}\label{split} 
with truncation errors of order $\tau ^3$. Furthermore, Yoshida has shown \cite{yoshida} that the symplectic integrators for $n\geq 2$ and $n$ even are constructed by the composition of 2nd order ones. More specifically, he proved that the constants satisfying a 4th order symplectic scheme:
\begin{eqnarray}\label{4si}
S_4(\tau (A+B))=S_2(\kappa_1 \tau (A+B)) \cdot S_2(\kappa_2 \tau (A+B)) \cdot 
S_2(\kappa_1 \tau (A+B))+ {\cal O}(\tau ^5) ~~,
\end{eqnarray} 
are the solutions of the algebraic equations: $2\kappa_1 +\kappa_2 =1 $, $2\kappa_1^3 +\kappa_2^3 =0 $.

\subsection{The tangent map method} 
\label{tmm}

As already mentioned in Section \ref{VE}, the variational equations of (\ref{hamlri})
constitute a time-dependent Hamiltonian system and therefore admit a symplectic integration 
scheme (see \cite{Skokos2010} for more details). The extended Hamiltonian (\ref{hamvar2})
is defined on the tangent space of the phase space solutions and consequently the operator $ L_H$ acts on the extended position momentum vector  $u=( x, p, \delta x, \delta p )$.
In particular, $ L_H$ decomposes again into the kinetic-potential operators: 
\begin{eqnarray}
e^{\tau A/2} u: \left\{
\begin{array} {lll}
\tilde{x} _k =\frac{\partial  H}{\partial p_k}  \cdot \frac{\tau} {2}+ x_k, ~~~~~~\nonumber\\
\tilde{p}_k = p_k\nonumber\\
\tilde{\delta x}_k = p_k  \cdot \frac{\tau} {2} + \delta x_k \nonumber\\
\tilde{\delta p}_k = \delta p_k  
\end{array}
 \right. ~~,
\end{eqnarray}
and 
\begin{eqnarray}
e^{\tau B} u: \left\{
\begin{array} {lll}
\tilde{x}_k = x_k,\nonumber\\
\tilde{p}_k = -\frac{\partial  H}{\partial x_k} \cdot  \tau + p_k \nonumber\\
\tilde{\delta x}_k = {\delta x}_k \nonumber\\
 \tilde{\delta p}_k = -\left( \sum_{j=1}^{N} \frac{\partial^2  H}{\partial x_k \partial x_j } \delta x_j \right) 
\cdot \tau + {\delta p}_k 
\end{array}
 \right. ~~,
\end{eqnarray}
through which we can now proceed to solve the variational equations of the problem.

\section{Applications to two important models}
\label{NR}

\subsection{Planar rotators and the mean field model}\label{NR1}
The 1D XY model is a chain of $N$ planar rotators, describing a system of spins coupled by nearest neighbor interactions. The implementation of LRI to this model has already been studied in various interesting papers \cite{latora,AnteneodoTsallis1998,CirtoAssisTsallis2013,Ruffo95,Dauxois2002} in connection with its statistical and dynamical properties.
The Hamiltonian of this system reads:
\begin{eqnarray}\label{prham}
{\cal H}=\frac{1}{2}\sum_{i} p_i^2 + \frac{1 }{2\tilde{N}}\sum_{i,j} \frac{1 - \cos 
(\vartheta_i - \vartheta_j )}{r_{i,j}^{\alpha }}
\end{eqnarray}
where for $\alpha \rightarrow \infty $ the classical nearest-neighbor rotators are recovered,
while for $ \alpha =0$ (\ref{prham}) reduces to the so--called HMF model. Regarding this model, Latora et al. \cite{latora} studied the $ \alpha =0$ case, while Anteneodo and Tsallis \cite{AnteneodoTsallis1998} explored the more general $ \alpha \geq 0$ system.

The variational equations for this model are easily derived from (\ref{prham}):
\begin{eqnarray}\label{vareq}
\dot{\vartheta }_i &=& p_i,\nonumber\\
\dot{p}_i &=& -\frac{1}{\tilde{N}} \sum_{j} \frac{\sin(\vartheta_i - \vartheta_j)}{r_{i,j}^{\alpha }}
\nonumber\\
\dot{\delta \vartheta }_i &=& {\delta p}_i \nonumber\\
\dot{\delta p}_i &=&  \sum_{j} a_{i,j} \delta \vartheta _j,~~~ 
\end{eqnarray}
where
\begin{eqnarray}
 a_{i,j}=  \left\{
 \begin{array} {ll}
-\frac{1}{\tilde{N}} \sum_{k}\frac{ \cos (\vartheta_i - \vartheta_k )}
{r_{i,k}^{\alpha }},~~~ {if}~~i=j
\nonumber\\
\frac{1}{\tilde{N}} \frac{ \cos (\vartheta_i - \vartheta_j )}
{r_{i,j}^{\alpha }}, ~~~~~~~~~~~{if}~~i\ne j
\end{array}
\right.
\end{eqnarray}

The system (\ref{vareq}) is used for the calculation of the maximal Lyapunov exponent $\lambda $. After a careful study of the behavior of the Lyapunov exponents in the HMF model, Latora et al. \cite{latora} established how they depend on the specific energy $\varepsilon = U(N)/N$ and the system size. More specifically, in a graph of $\lambda$ versus the specific energy, all data show a peak around $\varepsilon =0.67$, which weakly depends on the system size. Furthermore, $\lambda$ exhibits a decreasing level of chaos with increasing $N$ as it decays 
as $\lambda \sim N^{-1/3}$. In the same spirit, Anteneodo et al. in \cite{AnteneodoTsallis1998} also demonstrated a power law decay of $\lambda $ vs. $N$ for the system (\ref{prham}) and any $\alpha < 1$. 

{\  As we discuss below, the FPU model under LRI exhibits a power-law decay of the maximal Lyapunov exponent with increasing
$N$ similar to the coupled rotators system. However, remarkably enough in the FPU case $\lambda $ does not decay as the specific energy increases, but continues to grow as a power law.     }

\subsection{The FPU-$\beta $ 1D particle chain with LRI}

\begin{figure}
\centering
\includegraphics[width=0.45\linewidth]{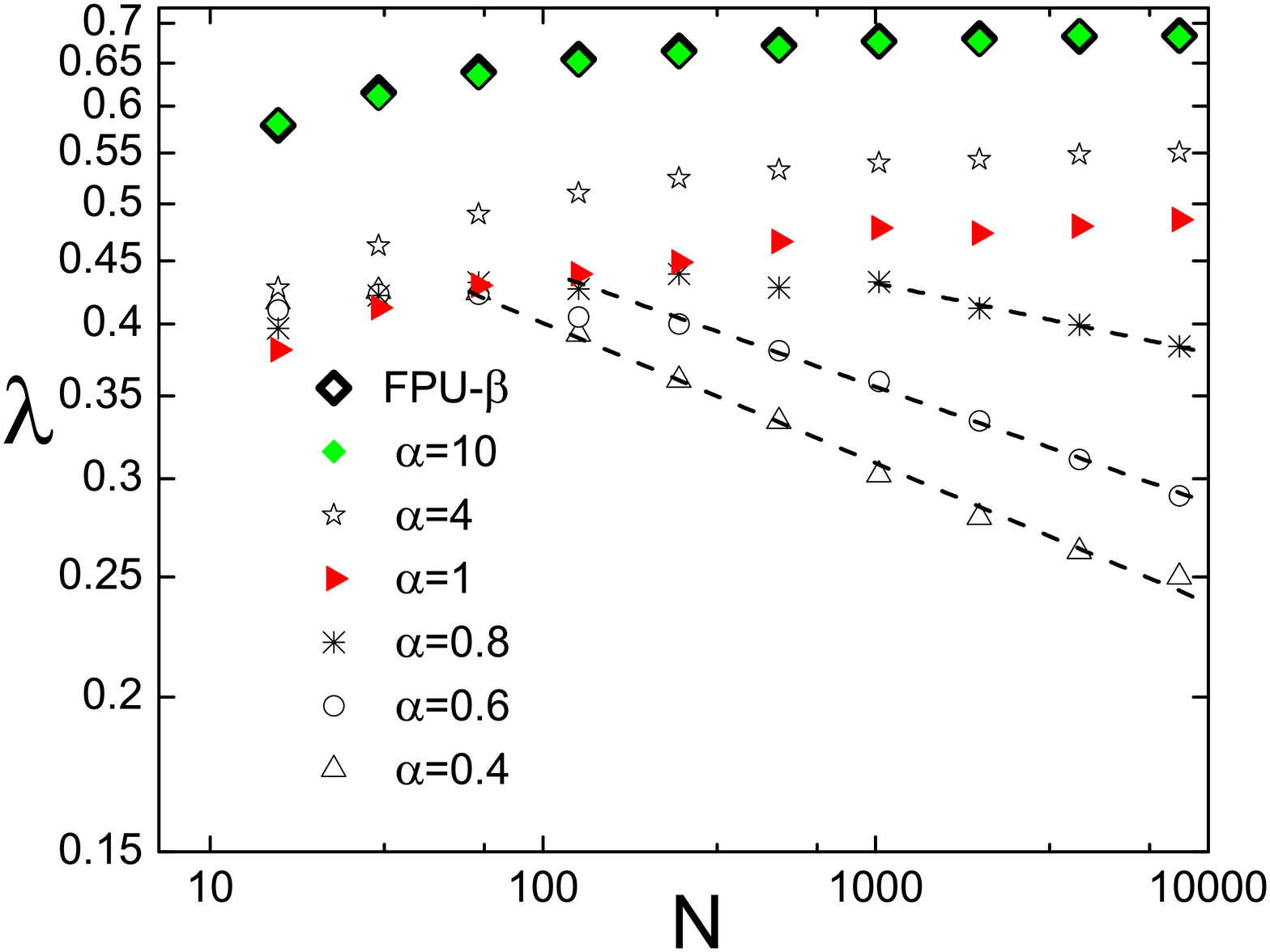}
\includegraphics[width=0.45\linewidth]{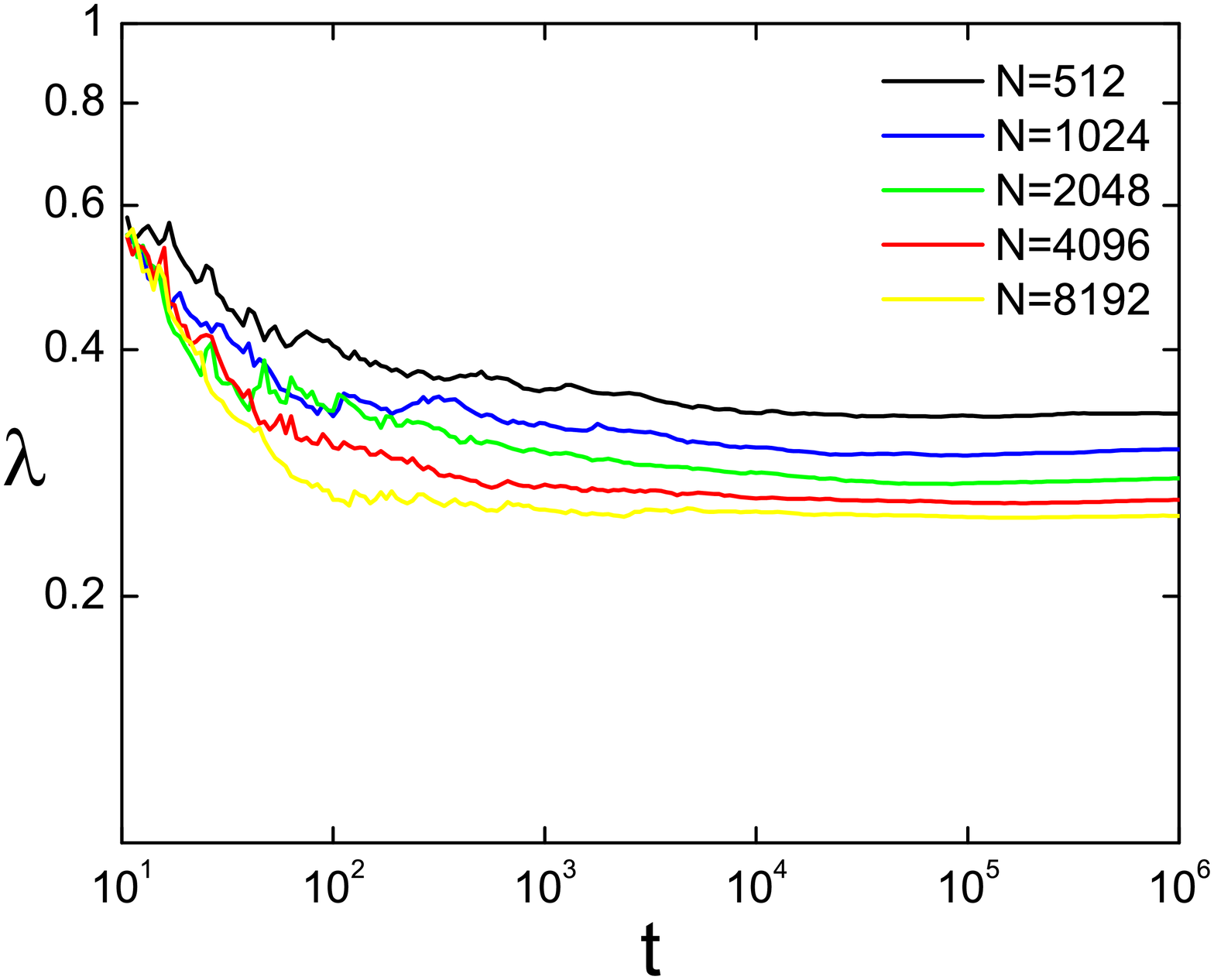}
\caption{ (a) The dependence of the maximal Lyapunov exponent on $N$, calculated at $t=10^6$ for various ranges of nonlinear interaction $\alpha $ ($\varepsilon =9$, $\beta =10$ and FPU-$\beta$ means $\alpha=\infty$). (b) Evolution of the maximal Lyapunov exponent for $\alpha =0.4$, $\varepsilon =9$ and $\beta =10$ for various system sizes $N$.}\label{LEmax}
\end{figure}

In the case of the FPU-$\beta $ chain, LRI can be implemented in various ways since the system has a potential with two distinct nearest neighbor parts, a quadratic and a quartic one \cite{CTB} 
\begin{equation}\label{fpuhamNN}
{\cal H}=\frac{1}{2}\sum_{n=1}^{N} p_n^2 + \frac{1}{2}\sum_{n=1}^N (x_{n+1}-x_n)^2 
+ \frac{\beta}{4} \sum_{n=1}^{N}(x_{n+1}-x_n)^4=U(N) \;\;\;(\beta>0),
\end{equation}
Clearly, the most general approach would be to add LRI with different $\alpha$ exponents to each part of the potential as follows:
\begin{eqnarray}\label{fpuham}
{\cal H}=\frac{1}{2}\sum_{i} p_i^2 + \frac{1}{4\tilde{N}_1}\sum_{i,j} \frac{(x_i - x_j)^2}{r_{i,j}^{\alpha_1 }}
 + \frac{\beta }{8\tilde{N}_2}\sum_{i,j} \frac{(x_i - x_j)^4}{r_{i,j}^{\alpha_2 }}=U(N)
\end{eqnarray}
providing thus the system with two independent ranges, an $\alpha_1 $ controlling 
the quadratic part of the potential and an $\alpha_2 $ controlling the quartic part. One can then focus on the computation of the maximal Lyapunov exponent $\lambda $, as $N$ increases at fixed specific energy $\varepsilon = U(N)/N$. 
To this end we need to integrate the variational equations:
\begin{eqnarray}
\dot{x}_i &=& p_i,\nonumber\\
\dot{p}_i &=& -\frac{1}{\tilde{N}_1} \sum_{j} \frac{x_i - x_j}{r_{i,j}^{\alpha_1 }}
- \frac{\beta}{\tilde{N}_2} \sum_{j} \frac{(x_i - x_j)^3}{r_{i,j}^{\alpha_2 }}  \nonumber\\
\dot{\delta x}_i &=& {\delta p}_i \nonumber\\
\dot{\delta p}_i &=&  \sum_{j} a_{i,j} \delta x_j,~~~ 
\label{LRIFPU}
\end{eqnarray}
with
\begin{eqnarray}
 a_{i,j}= \left\{
 \begin{array} {ll}
-\frac{\mu_i}{\tilde{N}_1} - \frac{3 \beta}{\tilde{N}_2} \sum_{k}
\frac{(x_i - x_k)^2}{r_{i,k}^{\alpha _2 }},~~~ {if}~~i=j\nonumber\\
\frac{\mu_i}{\tilde{N}_1} \cdot \frac{1}{r_{i,j}^{\alpha _1} } + \frac{3 \beta}{\tilde{N}_2} 
\frac{(x_i - x_j)^2}{r_{i,j}^{\alpha _2 }}, ~~~~{if}~~i\ne j
\end{array}
\right. 
\end{eqnarray}
using the tangent map method of Section \ref{tmm}, with $\mu_i= \sum_{j}r_{i,j}^{-\alpha_1 }$. 
The specific symplectic scheme we have employed is the Yoshida 4th order exponential splitting (\ref{4si}),
while the method for calculating the Lyapunov exponent is based on Benettin et al. \cite{benettin76}.
In Fig.\ref{LEmax} we plot the results of the model with LRI imposed only on the quartic potential, 
i.e. $\alpha _1\rightarrow \infty $ and $\alpha := \alpha _2 \geq 0$,
so that only the long range interactions involved in (\ref{LRIFPU}) are nonlinear. {  Furthermore, 
in all computations of this section we apply fixed boundary conditions, 
while periodic boundaries give very similar results. The initial conditions correspond to all positions
equal to zero and momenta drawn randomly from a uniform distribution.   
}

The evolution of the maximal Lyapunov exponent $\lambda $ is displayed in Fig. \ref{LEmax}(a) versus the system size $N$ in double logarithmic scale for several nonlinear ranges $\alpha$ and energies that vary proportionally to $N$, keeping the
specific energy $\varepsilon = 9$ fixed. As pointed out in previous studies, we also find here that for $\alpha < 1$ the
Lyapunov exponent decays as $N$ increases by a power-law of the form $\lambda (N) \sim N^{-\kappa }$, $\kappa >0$ implying
thus that the system tends to become less chaotic as the degrees of freedom grow. { Instead for $\alpha \geq 1$ the Lyapunov exponents tend to stabilize at a constant value. The higher is the $\alpha $ value, the larger the $\lambda $. Finally, for 
$\alpha =10$ we practically recover the classical FPU-$\beta $ model and as can be seen in Fig.\ref{LEmax}(a) the symbols of these two cases superpose perfectly.}
Plotting then in Fig. \ref{LEmax}(b) $\lambda $ as a function of time for $N=512,1024,2048,4096,8192$, $\alpha_2 =0.4$ and $\beta =10$ we compute these decreasing values of $\lambda $ at $t=10^6$ where they appear to stabilize for times greater than $10^4$.  

The challenging question, therefore, is whether $\lambda$ continues to decrease in the thermodynamic limit, where both $N$ and $U(N)$ increase indefinitely. This would be quite surprising since it would imply that the system would show signs of integrability in that limit! A more plausible alternative, of course, is that the dynamics tends to an ``edge of chaos'' state where it continues to be chaotic, albeit very weakly so, with a maximal Lyapunov exponent equal to zero. Since our computational capacity does not allow us to shed more light on this question, we shall have to postpone its investigation to a future analysis.

\begin{figure}
\centering
\includegraphics[width=0.45\linewidth]{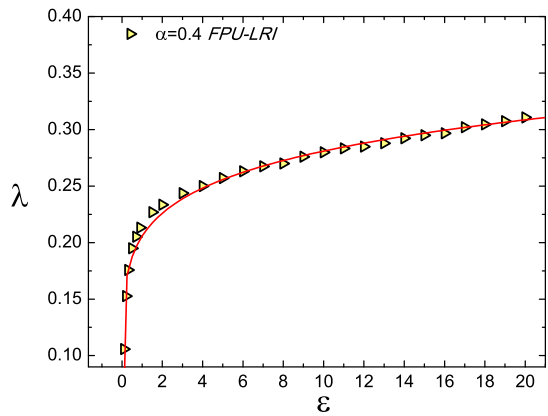}
\caption{ Numerical calculation of the maximal Lyapunov exponent as a function of the specific energy
for the system with $N=2048$, $\alpha =0.4$ and $\beta=10$. The red line corresponds to 
$\lambda = 0.205 \varepsilon ^{0.135}$. \label{LEvenergy}}
\end{figure}

We have also studied the dependence of the maximal Lyapunov exponent on the specific energy of the system. Recall that in subsection \ref{NR1} we mentioned that $\lambda $ has been found in the planar rotators model to peak at a specific $\varepsilon $-value. To probing further this possibility, we studied in Fig.\ref{LEvenergy} the behavior of $\lambda $ versus $\varepsilon $ of the FPU--$\beta$ model with $\alpha =0.4$. In contrast to the case of planar rotators, we observe that in the FPU system there is a monotonic increase of the maximal Lyapunov exponent with specific energy. { 
In particular, $\lambda $ increases with the specific energy as $\lambda = 0.205 \varepsilon ^{0.135}$.  The same behavior is conjectured for all $\alpha $ values, since similar results have emerged for the classical FPU--$\beta$ model as well. Nevertheless, the precise scaling laws depending on $\alpha $ will be studied in a future publication.}

\begin{figure}
\centering
\includegraphics[width=0.45\linewidth]{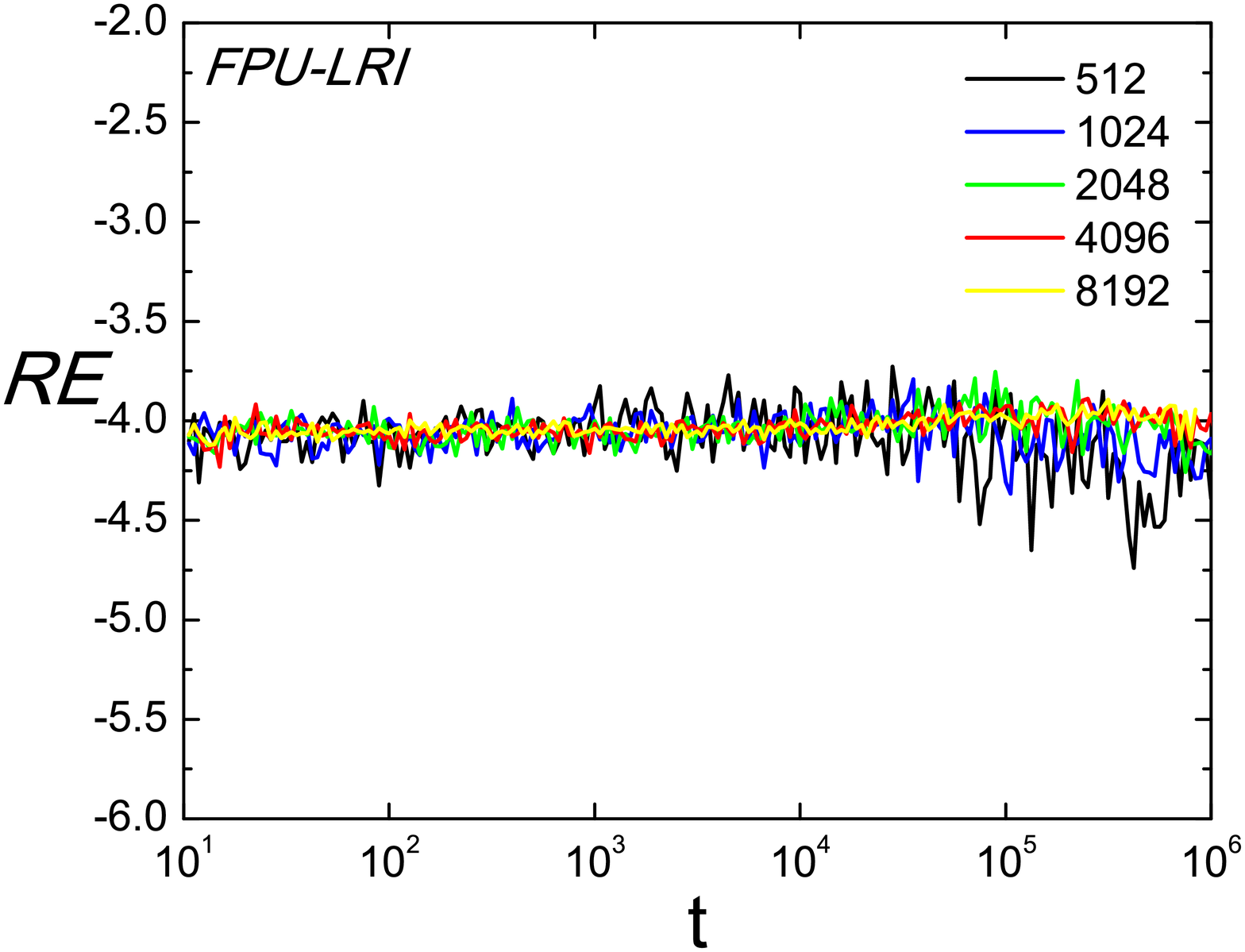}
\includegraphics[width=0.45\linewidth]{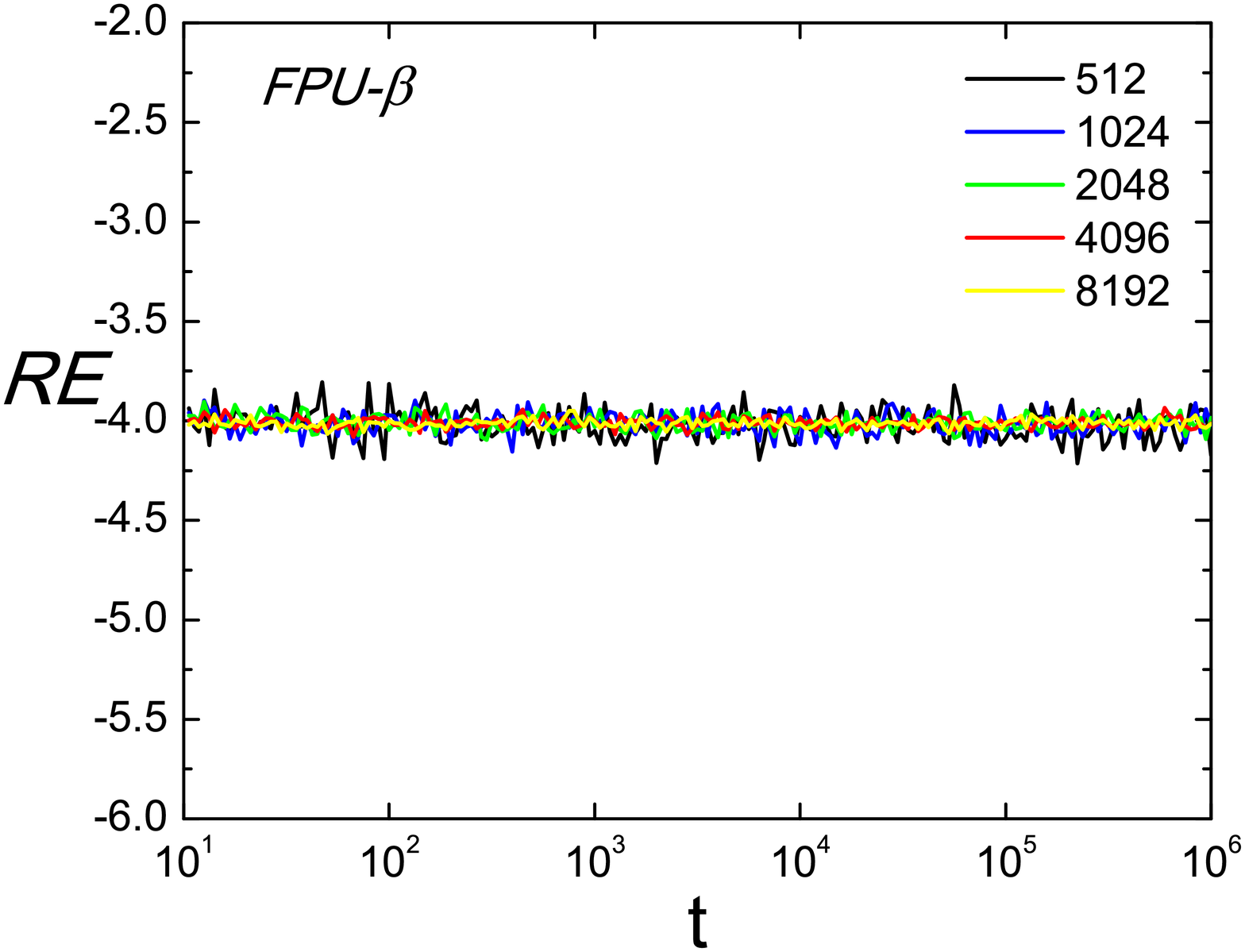}
\caption{The relative errors of: (a)  the Hamiltonian FPU-$\beta $ with LRI and (b) 
the classical FPU-$\beta $ Hamiltonian. Different 
system sizes have been considered and same time step $\tau =0.02$.} 
\label{error}
\end{figure}

Finally, as a test for the accuracy of our calculations, we compute the relative errors 
$$
{ RE} =\log \Big| \frac{E(t)-E(0)}{E(0)}  \Big| ~~,
$$
where $E(t)$ represents the total energy of the solutions as a function of time. Fig. \ref{error}(a) shows the relative errors of the Hamiltonian (\ref{fpuham}) for the LRI data of Fig.\ref{LEmax}(b), while Fig.\ref{error}(b) refers to relative errors in the FPU-$\beta $ model with only nearest neighbor interactions. Our results verify in each case that the integrated Hamiltonian $\tilde{H}$ remains within the same accuracy (nearly 4 significant digits) close to the desired Hamiltonian $H$ for a time step $\tau =0.02$. 
In addition, the energy fluctuations become smaller and smaller in both cases when $N$ increases. {  This is interesting because it implies that $RE$ fluctuations are independent of the behavior of
the maximal Lyapunov exponent, which is known to increase as a function of $N$ for the classical nearest-neighbor FPU system 
(Fig.\ref{error}(b)), while it decreases for the FPU with LRI (Fig.\ref{error}(a)). }
Nevertheless, we have no explanation why the energy fluctations are more pronounced
in systems with long range interactions compared to nearest neighbor systems like FPU.

\section{Conclusions}
\label{concl}

The study of the range of interactions in dynamical systems describing large numbers of particles is of great interest in many branches of science. In statistical physics, for instance, where the particles are modeled by nonlinear mechanical oscillators it is very important to understand the effect of mutual interactions on the dynamics and statistics of 1--dimensional Hamiltonian lattices in the so--called thermodynamic limit, where the number of particles $N$ and their total energy $E$ tend to infinity with $\varepsilon=E/N$ constant. For example, in a system of $N$ coupled planar rotators that interact via long--range forces critical regions have been found in connection with long-living quasi--stationary states. On the other hand, such states were also studied in a generalized mean field system, where mutual interactions decay with distance according to $1/r^{\alpha }$, and it was found that the largest Lyapunov exponent decays as a power-law with $N$. 

In the present paper we have used advanced numerical techniques to extend previous studies of nearest-neighbor Hamiltonian lattices to analogous models involving long range interactions. To achieve this, we have employed symplectic integrator schemes and have computed the tangent dynamics needed to calculate the largest Lyapunov exponent of two different Hamiltonian systems: the mean field model of planar rotors and the FPU-$\beta $ chain, both involving interactions whose range is modulated by the factor $1/r^{\alpha }$. We have thus been able to identify in these systems a continuous transition from strongly to weakly chaotic dynamics in the thermodynamic limit, as the parameter $\alpha $ varies from infinity to zero.

We thus conclude that $\alpha = 1$ appears to constitute a critical crossover value where a qualitative change occurs in the dynamics and statistics of such systems. As a chaotic index we have focused on the largest Lyapunov exponent $\lambda$ calculated efficiently and accurately using Yoshida's 4th order symplectic integrator scheme. For both models treated in this paper we found that the Lyapunov exponent gives the same results: (i) For $\alpha \geq 1$, $\lambda$ tends to stabilize at a positive value as $N$ increases, (ii) for $\alpha < 1$ $\lambda$ {\emph decreases} with system size as $N^{-\kappa(\alpha )}$, for some positive constant $\kappa(\alpha)$ that depends on $\alpha $. On the other hand, unlike the mean field rotator model, the $\lambda$ of the FPU-$\beta $ chain increases monotonically as the specific energy $\varepsilon=E/N$ is increased.
 
It is certainly somewhat intriguing that when long range interactions are applied {\emph only} to the harmonic part of the potential no crossover is observed at $\alpha = 1$, as the dynamics remains strongly chaotic and the statistics is always of the Boltzmann--Gibbs type \cite{CTB}. On the other hand, all the studies carried out so far regarding long range interactions have been performed on Hamiltonian systems involving only mutual interactions and no on site potential. It would, therefore, be very interesting to study these cases also in a future publication to complete our understanding of the effect of the range of interactions in 1--dimensional Hamiltonian lattices.

\paragraph{Acknowledgments}
This research has been co-financed by the European Union (European Social Fund - ESF) and Greek national funds through the Operational Program ``Education and Lifelong Learning'' of the National Strategic Reference Framework (NSRF) - Research Funding Program: THALES - Investing in knowledge society through the European Social Fund. Computer simulations were performed in the facilities offered by the High Performance Computing Systems and Distance Learning Lab (HPCS-DL Lab),  Technological Educational Institute of Western Greece.

\bibliographystyle{elsarticle-num}

\end{document}